\documentclass[12pt]{article}
\usepackage{graphicx} % Required for inserting images
\usepackage{bm}
\usepackage{authblk}
\usepackage{amsmath}
\usepackage{graphicx}
\usepackage{xcolor}
\usepackage{abstract}

\usepackage[paper=letterpaper]{geometry}
\makeatletter
\renewcommand{\maketitle}{\bgroup\setlength{\parindent}{0pt}
\begin{flushleft}
  \textbf{\@title}

  \@author
\end{flushleft}\egroup
}
\let\saved@includegraphics\includegraphics
\AtBeginDocument{\let\includegraphics\saved@includegraphics}
\renewenvironment*{figure}{\@float{figure}}{\end@float}
\makeatother
\usepackage[labelfont=bf]{caption}

\usepackage[style=nature,sorting=none,backend=biber,maxbibnames=99]{biblatex}
\AtEveryBibitem{%
  \ifentrytype{incollection}{%
    \DeclareFieldFormat{title}{#1}%
  }{}%
}

\DeclareBibliographyDriver{incollection}{%
  \printnames{author}%
  \adddotspace%
  \printfield{title}%
  \newunit
  \printfield{booktitle}%
  \newunit
  \printfield{volume}%
  \newunit
  \printfield{number}%
  \newunit
  \printfield{pages}%
  \newunit
  \printfield{year}%
}
\usepackage{soul}
\usepackage{comment}
\title{Control of out-of-plane anti-damping spin torque with a canted ferromagnetic spin source\footnote{$^\ast$  E-mail: xh384@cornell.edu, dcr14@cornell.edu}}

\author[1,11,$\ast$]{Xiaoxi Huang}
\author[1,11]{Daniel A. Pharis}
\author[2]{Hang Zhou}
\author[3,8]{Zishen Tian}
\author[1]{Thow Min Jerald Cham}
\author[4]{Kyoungjun Lee}
\author[5]{Yilin Evan Li}
\author[6]{Chaoyang Wang}
\author[1]{Yuhan Liang}
\author[1]{Maciej Olszewski}
\author[6]{Di Yi}
\author[4]{Chang-Beom Eom}
\author[5,7,10]{Darrell G. Schlom}
\author[3,8,9]{Lane W. Martin}
\author[2]{Ding-Fu Shao}
\author[1,7,$\ast$]{Daniel C. Ralph}

\affil[1]{Department of Physics, Cornell University, Ithaca, NY 14853, USA}
\affil[2]{Key Laboratory of Materials Physics, Institute of Solid-State Physics, HFIPS, Chinese Academy of Sciences, Hefei 230031, China}
\affil[3]{Department of Materials Science and Engineering, University of California, Berkeley, CA 94720, USA}
\affil[4]{Department of Materials Science and Engineering, University of Wisconsin-Madison, Wisconsin 53706, USA}
\affil[5]{Department of Materials Science and Engineering, Cornell University, Ithaca, NY 14853, USA}
\affil[6]{State Key Laboratory of New Ceramics and Fine Processing, School of Materials Science and Engineering, Tsinghua University, Beijing 100084, China}
\affil[7]{Kavli Institute at Cornell for Nanoscale Science, Ithaca, NY 14853, USA}
\affil[8]{Rice Advanced Materials Institute, Rice University, Houston, TX 77005, USA}
\affil[9]{Department of Materials Science and NanoEngineering, Chemistry, and Physics and Astonomy, Rice University, Houston, TX 77005, USA}
\affil[10]{Leibniz-Institut für Kristallzüchtung, Max-Born-Str. 2, 12489 Berlin, Germany}
\affil[11]{These authors contributed equally: Xiaoxi Huang, Daniel A. Pharis}
\begin{document}
      \maketitle
      \vspace{10mm}
    %\textbf{Spin current with out-of-plane spin component plays a pivotal role to achieve next-generation high-density energy-efficient magnetic memory and logic devices; and symmetry is the key to determining the orientations of spin polarization borne by a spin current. The reduction of symmetry has been predominantly realized via lowering crystal symmetry that inevitably limits the abundance of material candidates. Here we report the experimental observation of electric-field induced out-of-plane spin component borne by vertically flowing spin currents in a ferromagnet-SrRuO$_3$, where symmetry breaking is achieved by the canting of magnetic moment. A maximum out-of-plane damping-like spin torque efficiency and spin Hall conductivity of $0.014\pm0.0045$ and $\sigma_{SH}^Z \sim$$12\pm3.8$$\times$10$^3$($\hbar$/2e)($\Omega^{-1}$m$^{-1}$) are detected at $90$ K. %Furthermore, this out-of-plane spin component can be tuned by the orientation of magnetic moment, revealing another advantage associated with unconventional torque generated by magnetic moment-orientability.
    %Our observations introduce a new strategy to effectively break system symmetry and achieve efficient generation of out-of-plane damping-like torque.}\\

    \textbf{
    To achieve efficient anti-damping switching of nanoscale magnetic memories with perpendicular magnetic anisotropy using spin-orbit torque requires that the anti-damping spin-orbit torque have a strong out-of-plane component.  The spin anomalous Hall effect and the planar Hall effect spin current produced by a ferromagnetic layer are candidate mechanisms for producing such an out-of-plane anti-damping torque, but both require that the magnetic moment of the spin source layer be canted partly out of the sample plane at zero applied magnetic field.  Here we demonstrate such a canted configuration for a ferromagnetic SrRuO$_3$ layer and we characterize all vector components of the torque that it produces, including non-zero out-of-plane anti-damping torques. We verify that the out-of-plane spin component can be tuned by the orientation of magnetic moment, with significant contributions from both the spin anomalous Hall effect and the planar Hall effect spin current.}
    \\

High-density magnetic random access memory (MRAM), which for thermal stability must be based on magnetic layers with perpendicular magnetic anisotropy (PMA), is attractive for its natural non-volatility and excellent endurance as compared to competing technologies \cite{bhatti2017spintronics,dieny2017perpendicular}.  Anti-damping spin-orbit torque exerted by spin currents generated within materials having strong spin-orbit interactions is a leading candidate for enhancing the write efficiency in such memories to make them competitive for energy-constrained applications \cite{mihai2010current,miron2011perpendicular,liu2011spin,liu2012spin,mellnik2014spin,fan2014magnetization,han2017room,wang2017room,nan2019anisotropic,huang2021novel}.  Unfortunately, spin-source materials with high symmetry can only produce anti-damping spin-orbit torque with orientation within the sample plane, which means that they are not capable of driving efficient anti-damping switching of nanoscale PMA magnets \cite{garello2013symmetry,zhang2015spin}.  Efficient switching of nanoscale PMA magnets by anti-damping spin-orbit-torque requires spin-source materials in which the symmetry constraints are relaxed either by a low-symmetry crystal structure \cite{macneill2017control,liu2021symmetry,dc2023observation,liu2023field,patton2023symmetry} or magnetic order \cite{taniguchi2015spin,vzelezny2017spin,iihama2018spin,nan2020controlling,chen2021observation,bose2022tilted,chen2025pure}, to allow a component of out-of-plane anti-damping torque.  Modeling suggests that an out-of-plane damping-like torque efficiency $\xi^Z_{DL} \approx$ 0.05 is necessary for spin-orbit-torque MRAM to achieve energy efficiency better than conventional spin-transfer-torque MRAM for 40 nm devices, while $\xi^Z_{DL} \approx$ 0.30 would allow better energy efficiency even than volatile SRAM with a much more compact circuit geometry \cite{liao2020spin}.  Research on developing spin-source materials for out-of-plane anti-damping torque has so far fallen short of these thresholds, except for perhaps the low-symmetry van der Waals semimetal TaIrTe$_4$, for which recent studies report conflicting results \cite{liu2023field,zhangTaIrTe4,bainsla2024large}.  Nonetheless, generating out-of-plane anti-damping torque with TaIrTe$_4$ requires domain-free flakes completely free of monolayer steps, so TaIrTe$_4$ is unlikely to be a solution for a manufacturable technology. \\

Here we pursue the strategy of generating an out-of-plane anti-damping torque with a ferromagnetic spin source layer in which the orientation of the ferromagnetic moment provides the required symmetry breaking.  Several mechanisms have been proposed and demonstrated with initial experiments for how ferromagnets may be used as a source of spin current with a spin orientation that is controllable by changing the magnetization direction \cite{amin2019intrinsic,davidson2020perspectives}:  e.g., via the spin anomalous Hall effect (AHE) \cite{taniguchi2015spin,iihama2018spin,seki2019large,montoya2025anomalous}, the planar Hall effect (PHE) spin current \cite{taniguchi2015spin,safranski2019spin,safranski2020planar}, or spin rotation or spin swapping at a magnetic interface \cite{saidaoui2016spin,humphries2017observation,baek2018spin,amin2018interface,yang2024field}. Initial experiments suggest that the spin AHE and the PHE spin current are particularly promising.  To generate a vertically flowing spin current with an out-of-plane spin component via either of these mechanisms requires that the magnetization of the spin-source layer be canted partially out of the sample plane (i.e., neither fully in the sample plane nor fully perpendicular to the plane):  the out-of-plane anti-damping torque due to the spin AHE is predicted to be proportional to $m_Y m_Z$ and that due to the PHE spin-current mechanism is predicted to be proportional to $m_X m^2_Z$ \cite{taniguchi2015spin}.  (Here $\hat{m}$ is the magnetic orientation, $\hat{X}$ is the direction of the in-plane applied charge current, and $\hat{Z}$ is perpendicular to the sample plane.)  The strength of the underlying physical mechanisms for both the spin AHE and PHE spin currents have been measured previously using samples with either in-plane magnetizations \cite{iihama2018spin,seki2019large} or with magnets canted using large saturating magnetic fields \cite{safranski2019spin,safranski2020planar,montoya2025anomalous}, with efficiencies as large as $\xi^Y_{DL}=$ 0.25 for the in-plane torque efficiency generated by in-plane-magnetized \textit{L}1$_0$ FePt \cite{seki2019large} via the spin AHE.  
%The in-plane torque efficiency for the spin anomalous Hall effect has been measured to be 0.25 for in-plane-magnetized L1$_0$ FePt \cite{iihama2018spin} and 0.14 for in-plane Co$_{60}$Fe$_{20}$B$_{20}$ {\color{blue}[Please re-cite T.\ Seki, S.\ Iihama, T.\ Taniguchi, and K.\ Takanashi, ``Large spin anomalous Hall effect in L10-FePt: Symmetry and magnetization switching,” Phys.\ Rev.\ B 100, 144427 (2019)}.  The effective torque efficiency of the planar spin Hall currents emitted by a Co/Ni superlattice and acting on CoFeB is reported as 0.05 \cite{safranski2020planar}.  Both the planar Hall spin current emitted from a Co/Ni superlattice and the spin anomalous Hall effect from Ni$_{90}$Fe$_{10}$ nanowires are sufficiently strong to fully compensate the magnetic damping and excite magnetic auto-oscillations \cite{safranski2019spin} {\color{blue}[please re-cite E.\ A.\ Montoya, X.\ Pei, and I.\ N.\ Krivorotov, ``Anomalous Hall spin current drives self-generated spin–orbit torque in a ferromagnet,” Nat.\ Nanotech.\ https://doi.org/10.1038/s41565-024-01819-7 (2025)]}.  
The challenge with using these mechanisms to generate out-of-plane anti-damping torque for practical applications is to stabilize a partially-canted magnetic configuration with zero applied magnetic field.  Thin-film samples governed entirely by shape anisotropy and first-order interfacial anisotropy have either purely in-plane or PMA configurations, so to have a stable partially-canted magnetic configuration requires use of magneto-crystalline anisotropy.   \\

    In a high-crystal-symmetry non-magnetic spin source material,
    %with spin degenerate Fermi surface (Fig.\ 1a), 
    an electric-field induced spin current that flows vertically can have a spin polarization only in an in-plane direction orthogonal to both the electric field and spin current (Fig.\ 1a). Fortunately, this constraint can be relaxed if the spin-source material has magnetic order. 
    %The spin-split Fermi surface of a ferromagnetic metal (Fig.\ 1c) allows for more possible spin orientations.
    The electric-field generated spin current produced by a ferromagnetic spin-source layer can have a part with the symmetry of the conventional in-plane spin-orbit torque that does not depend on the magnetization direction (Fig.\ 1a) plus a part that does depend on the magnetization direction \cite{amin2019intrinsic}. The geometries of the magnetization-dependent spin currents generated by the PHE and AHE \cite{taniguchi2015spin} are shown in Fig.\ 1b,c for two different orientations of the canted magnetization relative to an applied electric field, $\vec{E}$.  Both the PHE and AHE give rise to spin currents with their spin polarization aligned with magnetization of the spin-source layer, but with different flow directions: for the PHE spin current the flow is parallel to $\hat{m}(\hat{m}\cdot \vec{E})$ (Fig.\ 1b) and for the spin AHE the flow is along $\hat{m}\times \vec{E}$ (Fig.\ 1c). To achieve a vertically flowing spin current with an out-of-plane spin component, therefore, the PHE can contribute when the magnetization has a canted component in the $\hat{X}$-$\hat{Z}$ plane, and the AHE can similarly contribute when the magnetization has a canted component in the $\hat{Y}$-$\hat{Z}$ plane.  As a function of the angle $\psi$ of the applied electric field relative to the projection of the canted magnetization onto the sample plane, this means that the out-of-plane anti-damping torque should have a $\cos(\psi)$ dependence for the PHE spin current and a $\sin(\psi)$ dependence for the spin AHE.  These results can be generalized: based on a symmetry analysis any time-reversal-odd spin current contributing to an out-of-plane anti-damping torque should give a $\cos(\psi)$ dependence, and any time-reversal-even spin current should give a $\sin(\psi)$ dependence (see Supplementary Information Section 5).  \\

    To achieve a magnetic spin-source layer with a partially-canted magnetization at zero applied magnetic field, we employ SrRuO$_3$ as the spin source \cite{Ou2019,wei2021enhancement,zhou2021modulation} and deposit it on vicinal (001)-oriented SrTiO$_3$ substrates with a miscut angle around $0.1^\circ$ and a miscut direction close to $[010]$ (Methods). SrRuO$_3$ is an itinerant ferromagnet with a bulk Curie temperature $T_C \sim$ 160 K \cite{longo1968magnetic,schreiber2023enhanced} and an orthorhombic crystal structure with lattice parameters $a_o$ = 5.5670 $\AA$, $b_o$ = 5.5304 $\AA$, and $c_o$ = 7.8446 $\AA$ (the subscript $o$ denotes orthorhombic structure) \cite{jones1989structure}. In the bulk, the magnetic easy axis aligns with $[010]_o$, which we will call the $b$ axis \cite{gan1999lattice}, with a large magnetocrystalline anisotropy field, $\sim$ 12 T \cite{koster2012structure}.  When deposited on cubic, well-oriented (001)-SrTiO$_3$ substrates, SrRuO$_3$ thin films can grow epitaxially with six different domain configurations (see Supplementary Information Section 1).  Nevertheless, by nucleating growth from step edges on a vicinal substrate, a single domain can be made dominant \cite{gan1997control,jiang1998domain}.  With this method we achieve the predominant epitaxial configuration shown schematically in Fig.~\ref{fig:2}a, a (110)$_o$-oriented SrRuO$_3$ film with SrRuO$_3$ $[110]_o \parallel$ SrTiO$_3$ $[001]$, SrRuO$_3$ $[001]_o \parallel$ STO $[010]$, and SrRuO$_3$ $[\Bar{1}10]_o \parallel$ STO $[100]$, resulting in the bulk magnetic easy axis (i.e., the $b$ axis) canted uniformly away from the surface normal \cite{gan1999lattice}.  We verified this growth orientation using x-ray diffraction with reciprocal space mapping (Supplementary Information Section 1).  Unsuccessful growths, in which more than one domain type are present, can also be detected by measurements of anomalous Hall effect versus out-of-plane magnetic field, because the different domain types have different coercive fields (Extended Data Fig.~\ref{fig:efig_1}).  For convenience in the following, we will adopt pseudo-cubic (pc) notation for SrRuO$_3$ based on the SrTiO$_3$ crystal axes, so that $[100]_{pc}$, $[010]_{pc}$ and $[001]_{pc}$ are parallel to $[\Bar{1}10]_o$, $[001]_o$ and $[110]_o$.  \\
    
    In thin films of SrRuO$_3$ the orientation of the canted easy axis is tunable by oxygen octahedral rotation and  need not lie exactly along the $b$ axis \cite{lu2013control,lu2015strain,liu2019current}. We therefore performed angular-dependent magnetoresistance measurements to determine the orientation of the magnetic easy axis in our thin films \cite{lu2013control,liu2019current}. The device geometry is shown in Fig.~\ref{fig:2}b, where a longitudinal voltage ($V_{XX}$) is measured while rotating an external magnetic field ($\vec{B}$) in the plane perpendicular to the applied current. This magnetoresistance measurement was performed on devices oriented at in-plane angles from $0^\circ$ to $90^\circ$ at every $15^\circ$ (Supplementary Information Section 2), to identify both the magnetic canting direction and canting angle. A representative angular-dependent magnetoresistance measurement is shown in Fig.~\ref{fig:2}c, where the jumps around $115^\circ$ ($295^\circ$) in resistance indicate the orientation of the magnetic hard axis (Supplementary Information Section 2). Since the easy-axis angle ($\theta_C$) is 90$^\circ$ from the hard axis, for the 20 nm SrRuO$_3$ film $\theta_C\sim 25^\circ$ away from $[001]_{pc}$  towards $[100]_{pc}$. The canting angle of SrRuO$_3$ of various thicknesses is summarized in Fig.~\ref{fig:2}d, consistent with previous work \cite{yoo2006diverse,liu2019current}.  \\
    %Besides the canting of magnetization, an implicit requirement for generating out-of-plane damping-like torque with magnetic moment is domain control, as contributions from different domains could potentially cancel each other out. It has been shown that twin domain formation in SrRuO$_3$ thin films can be diminished and even suppressed via a selection of the miscut angle and miscut direction of the vicinal SrTiO$_3$ (STO) substrates\cite{gan1997control,jiang1998domain}. Therefore, predominantly single-domained SrRuO$_3$ thin films shown in Fig.~\ref{fig:2}a can be obtained (Extended Data Fig. 1, Supplementary Information Section 1).\\ 
   
   For spin-torque studies, after the SrRuO$_3$ films were grown they were transferred through air to a separate vacuum system for deposition of a 5 nm thick Ni$_{80}$Fe$_{20}$ (Py) layer by magnetron sputtering on top of the SrRuO$_3$ (Methods). The Py layer has within-plane magnetic anisotropy. For thin SrRuO$_3$ samples ($5$ nm), where a single domain in the SrRuO$_3$ can be easily achieved (Extended Data Fig.~\ref{fig:efig_1}b), the spin-torque measurements were performed on the as-grown samples, while the thicker samples ($>5$ nm) in which a second type of domain becomes more significant (Extended Data Fig.~\ref{fig:efig_1}d) were field-cooled to liquid nitrogen temperature  with a magnetic field oriented along the $[100]_{pc}$ axis prior to the spin torque measurements. Scans of longitudinal resistance as a function of in-plane magnetic field at $80$ K along both the $[100]_{pc}$ and $[010]_{pc}$ are, to a good approximation, symmetric about zero magnetic field (Extended Data Fig.~\ref{fig:efig_2}), which indicates that any exchange bias between the SrRuO$_3$ and Py is weak.  \\

    We measured the direction and strength of current-induced spin torques in the SrRuO$_3$/Py samples using angle-dependent spin torque ferromagnetic resonance (ST-FMR) \cite{liu2011spin,fang2011spin,macneill2017control,bose2022tilted} (Methods). A schematic illustration of our ST-FMR device geometry is shown in Fig.~\ref{fig:3}a. The device consists of a SrRuO$_3$/Py  strip  10 $\mu$m $\times$ 40 $\mu$m in contact with ground-signal-ground electrodes made from Ti/Pt. Strips with different in-plane orientations relative to the crystal axes were fabricated to study the dependence of spin-orbit torque generation on the angle of applied electric field (or, equivalently, the angle of applied current). On each device with a distinct orientation, we apply a microwave signal with fixed frequency and sweep the external magnetic field $\vec{B}$ at an angle $\varphi$ to the applied electric-field direction (Fig.~\ref{fig:3}a), repeating the measurement on the same device for $\varphi$ from $0^\circ$ to $170^\circ$. An example ST-FMR spectrum at $\varphi = 45^\circ$ is shown in Fig.~\ref{fig:3}b, where the mixing voltage ($V_{mix}$) is fit to the sum of symmetric Lorentzian, $V_S = S(\frac{\Delta^2}{(B-B_0)^2+\Delta^2})$, and anti-symmetric Lorentzian, $V_A = A(\frac{(B-B_0)\Delta}{(B-B_0)^2+\Delta^2})$. Here, $S$ and $A$ are the amplitudes of the symmetric and anti-symmetric components of the mixing voltage, $\Delta$ is the linewidth of the resonance peak, and $B_0$ is the resonant magnetic field. A pronounced difference in the amplitudes of the anti-symmetric component of the mixing voltage at positive and negative fields is observed, indicating the existence of an unconventional torque in the out-of-plane direction. To unambiguously identify the directions of the torques, symmetry analysis is necessary, since the signals generated by different components of torque have different dependencies on the orientations of the external magnetic field. If we define the current flow direction as $\hat{X}$, the out-of-plane direction as $\hat{Z}$, and use polar coordinates to describe the magnetic field with the azimuthal angle $\varphi$ measured from the $\hat{X}$ direction, then the $\varphi$ dependence of the mixing voltages from X, Y, Z components of the torques are \cite{macneill2017control,bose2022tilted}:
    \begin{equation}
        V_A = A_{FL}^X \sin2\varphi \sin\varphi + A_{FL}^Y \sin2\varphi \cos\varphi + A_{DL}^Z \sin2\varphi 
        \label{V_A}
    \end{equation}
    \begin{equation}
        V_S = S_{DL}^X \sin2\varphi \sin\varphi + S_{DL}^Y \sin2\varphi \cos\varphi + S_{FL}^Z \sin2\varphi 
        \label{V_S}
    \end{equation}
    where $S_{DL}^X$, $S_{DL}^Y$, $A_{DL}^Z$ are the amplitudes of the mixing voltages contributed by the X, Y, Z components of the damping-like torque; and $A_{FL}^X$, $A_{FL}^Y$, $S_{FL}^Z$ are the amplitudes of the mixing voltages from the contributions of the X, Y, Z components of the field-like torque. %Since $A_{FL}^Y$ is mainly contributed by Oersted magnetic field produced by the microwave current in the SrRuO$_3$ layer, 
    Two representative angle-dependent ST-FMR results at 110 K are shown in Fig.~\ref{fig:3}c,d and Fig.~\ref{fig:3}e,f for a SrRuO$_3$ (5 nm)/Py (5 nm) sample with applied electric fields approximately parallel to the $[010]_{pc}$ and $[100]_{pc}$. For both orientations of $\vec{E}$, fits to both the anti-symmetric and symmetric mixing voltages require additional Z ($\sin2\varphi$) and X ($\sin2\varphi \sin\varphi$) components of the torques, in addition to the conventional Y component ($\sin2\varphi \cos\varphi$). %Since the magnetization of SrRuO$_3$ is canted away from $[001]_{pc}$ towards $[100]_{pc}$, the unconventional torques produced by electric field parallel to $[010]_{pc}$ are expected to be excited by AHE. 
    Comparing the ST-FMR results for two perpendicular electric-field directions, the anti-symmetric mixing signals corresponding to the Z component of torque have opposite signs (compare the red lines in Fig.~\ref{fig:3}c and Fig.~\ref{fig:3}e). Below we will ascribe this difference to separate contributions from the spin AHE and the PHE spin current.  \\ %This distinctly different anti-symmetric mixing voltage from Z component is attributed to the fact that when the electric field is applied along $[100]_{pc}$ direction, the out-of-plane damping-like torque is generated by PHE and spin current from AHE is prohibited in this geometry.\\

    We have confirmed the presence of the unconventional torque components using a second measurement technique, second harmonic Hall measurements. Angle-dependent second harmonic Hall voltage data for 100 K are shown in Extended Data Fig.~\ref{fig:efig_4}a and b for the same SrRuO$_3$ (5 nm)/Py (5 nm) sample with applied electric fields approximately parallel to the $[010]_{pc}$ and $[100]_{pc}$. The Z component of the damping-like torques for electric-field along $[010]_{pc}$ and $[100]_{pc}$ have opposite signs, in agreement with the ST-FMR results. \\

    Based on the ST-FMR measurements, the damping-like torque efficiencies can be be calculated as \cite{karimeddiny2020transverse}
    \begin{equation}
    \xi_{DL}^Y = S_{DL}^Y\frac{eM_St_{FM}}{\hbar J_e}\frac{4\Delta}{I_{rf}R_{AMR}}\frac{2B_0+\mu_0M_{eff}}{\sqrt{B_0(B_0+\mu_0M_{eff})}}
    \label{Y_2}
\end{equation}
\begin{equation}
    \xi_{DL}^Z = A_{DL}^Z\frac{eM_St_{FM}}{\hbar J_e}\frac{4\Delta}{I_{rf}R_{AMR}}\frac{2B_0+\mu_0M_{eff}}{B_0+\mu_0M_{eff}}
    \label{Z_2}
\end{equation}
    where $M_s$ (A/m) is the saturation magnetization for 5 nm of Py, $t_{Py}$ is the thickness of the Py layer, $I_{rf}$ is the amplitude of the microwave current within the full bilayer, $J_e$ is the current density within the SrRuO$_3$, and $\mu_0M_{eff}$ (T) is the  effective field from the magnetic anisotropy.  We calibrate $I_{rf}$ by comparing the heating-induced change in resistance due to a microwave current to the heating-induced change of a d.c.\ current, and then $J_e$ is determined from a parallel circuit model using the measured resistivities of the Py and the SrRuO$_3$ (see Supplementary Information section 3).  For the data shown in Fig.~3, $I_{rf} = 2.03$ mA and $J_e = 9.50\times10^9$ A/m$^2$.
    For the case of $\vec{E}$ along the [010]$_{pc}$, the Y and Z components of the damping-like torque efficiencies estimated with Eq.~\ref{Y_2} and Eq.~\ref{Z_2} are $\xi^Y_{DL} =0.45 \pm 0.06$ and $\xi^Z_{DL} =0.009 \pm 0.001$. Given that the resistivity ($\rho$) of SrRuO$_3$ at 110 K is around 231 $\mu \Omega$cm, the spin Hall conductivities estimated using $\sigma_{DL} = (\hbar / 2e) \xi_{DL}/\rho$ for the Y and Z components are $\sigma^Y_{DL} =(2.0\pm 0.3)\times 10^5 (\hbar/2e)(\Omega^{-1}$m$^{-1}$) and $\sigma^Z_{DL} =(3.9\pm 0.4)\times 10^3 (\hbar/2e)(\Omega^{-1}$m$^{-1}$). For $\vec{E}$ along the [100]$_{pc}$, the damping-like torque efficiencies for Y and Z components are estimated to be $\xi^Y_{DL} =0.49\pm 0.07$ and $\xi^Z_{DL} =-0.010\pm 0.001$; and the corresponding spin Hall conductivities for Y and Z component are $\sigma^Y_{DL} =(2.1\pm 0.3)\times 10^5 (\hbar/2e)(\Omega^{-1}$m$^{-1}$) and $\sigma^Z_{DL}=(-4.3\pm 0.4)\times 10^3 (\hbar/2e)(\Omega^{-1}$m$^{-1}$).  \\
    %The damping-like torque efficiencies can be estimated (approximately) using:
    %\begin{equation}
    %    \xi_{DL}^Y = \frac{S_{DL}^Y}{A_{FL}^Y} \frac{e \mu_0 M_S t_{Py} d_{SRO}}{\hbar} \sqrt{1+\frac{\mu_0 M_{eff}}{B_0}}
    %    \label{Y}
   % \end{equation}
   % \begin{equation}
   %     \xi_{DL}^Z = \frac{A_{DL}^Z}{A_{FL}^Y} \frac{e \mu_0 M_S t_{Py} d_{SRO}}{\hbar} 
   %     \label{Z}
   % \end{equation}
    %where $Ms$ = 1000 kA/m is the saturation magnetization of 5 nm Py, $t_{Py}$ = 5 nm is the thickness of Py, $d_{SRO}$ is the thickness of SrRuO$_3$, and $\mu_0M_{eff}$ = $0.7$ T is the  effective field from the magnetic anisotropy.
   % These expressions assume that the torque associated with $A^Y_{FL}$ is due entirely to the Oersted magnetic field, with negligible contribution from a field-like spin-orbit torque.  To verify this assumption, a detailed discussion on the calculation of Oersted field contribution to $A_{FL}^Y$ based on microwave current calibration is shown in Supplementary Information Section 3, suggesting that the Oersted field contribution accounts for $59\%$ of the measured anti-symmetric component of the mixing voltage in the SrRuO$_3$ ($5$ nm)/Py ($5$ nm) sample. 

    To investigate the origins of out-of-plane anti-damping torque, the magnetic field angle-dependent ST-FMR measurements were repeated on devices with intermediate $\psi$ angles for the orientation of the applied electric field relative to the [100]$_{pc}$ crystal direction (Fig.~\ref{fig:4}a).  The out-of-plane anti-damping torque efficiency $\xi_{DL}^Z$ has the angular dependence shown in Fig.~\ref{fig:4}b. This behavior of $\xi_{DL}^Z$ is in excellent agreement with the predicted dependence for the sum of torques due to the spin AHE ($\propto \sin(\psi)$) and the PHE spin current ($\propto \cos(\psi)$): $\xi_{DL}^Z = \xi_{PHE}^Z \cos\psi + \xi_{AHE}^Z \sin\psi$ for fitting parameters $\xi_{PHE}^Z$ = -0.0070 and $\xi_{AHE}^Z$ = 0.0067 for the 5 nm thick SrRuO$_3$.\\
    
     %The dependence of the out-of-plane anti-damping torque on the thickness of the SrRuO$_3$ layer and on temperature are shown in Fig.~\ref{fig:4}. 
     In addition to data for the 5 nm device, Fig.~\ref{fig:4}b  also shows the measured out-of-plane anti-damping torque efficiencies $\xi^Z_{DL}$ per unit charge density as a function of the electric-field angle $\psi$ for 20 and 30 nm SrRuO$_3$ layer thicknesses. The temperature dependences for the peak out-of-plane anti-damping torque efficiencies for SrRuO$_3$ of 5 nm, 20 nm and 30 nm are shown in Fig.~\ref{fig:4}c.  We find that $\xi_{DL}^Z$ reaches a peak value at approximately 20 nm, with $\xi_{DL}^Z =$ 0.020 $\pm$0.002 at 90 K.  This peak torque efficiency is greater for 20 nm thick SrRuO$_3$ than for 30 nm despite the fact that the 30 nm thick layer has a comparable canting angle for the magnetization (see Fig.~\ref{fig:2}d). This difference in torque efficiency could be an indication, e.g., of competing contributions from the larger magnetization canting angle versus a less-perfect domain structure for the 30 nm films (Extended Data Fig.~\ref{fig:efig_1}), or a contribution from an interfacial torque mechanism and not just a bulk mechanism. \\
   %({\color{red}{The out-of-plane anti-damping spin torque efficiency $\xi^Z_{DL}$ per unit charge density within the SrRuO$_3$ decreases moderately as SrRuO$_3$ thickness is increased (Fig.~\ref{fig:4}b), suggesting contributions from both interfacial mechanisms and bulk spin AHE and PHE spin current mechanisms.}}) 
   
    The temperature dependence of $\xi^Z_{DL}$ for a SrRuO$_3$ (5 nm)/Py (5 nm) is shown in Fig.~\ref{fig:4}d for the electric-field angle $\psi =120^\circ$, near which the magnitude of $\xi^Z_{DL}$ is maximized. The magnitude of $\xi_{DL}^Z$ decreases with increasing temperature and approaches zero near 135 K, just below the bulk Curie temperature (and remains zero up to room temperature, see Extended Data Fig.~\ref{fig:efig_6}). (The Curie temperature ($T_C$) of the SrRuO$_3$ thin films of various thicknesses can be identified from the kink in the resistivity versus temperature curves (Fig.~\ref{fig:4}e), yielding values of $T_C$ between 130 K and 140 K.) Based on measurements at $\psi =$ 0$^\circ$ and 90$^\circ$, we can also separate the temperature dependence due to the PHE ($\propto \cos(\psi)$) and spin AHE ($\propto \sin(\psi)$) spin current mechanisms (Fig.~\ref{fig:4}f). The spin AHE contribution has very little temperature dependence between liquid-nitrogen temperature and 130 K, while the PHE spin current contribution varies strongly in the same temperature range.  We speculate that the time-even spin AHE contribution is due primarily to intrinsic spin-orbit coupling within the SrRuO$_3$ bandstructure and is therefore relatively temperature-independent well below the Curie temperature, while the PHE spin current is odd under time reversal and is therefore more likely to depend on extrinsic scattering and other processes which can vary with temperature.\\  %should depend on the temperature-dependent Fermi level {\color{red}{[It seems that when the Fermi surface changes with temperature, so does Berry curvature.]}} in SrRuO$_3$ \cite{wang2020controllable,sohn2021sign} and extrinsic scattering processes, making it more sensitive to changes with temperature. \\

    The out-of-plane anti-damping torque efficiency we find from SrRuO$_3$ is less than about half the value needed to directly drive efficient anti-damping switching of nanoscale magnetic samples with PMA, as indicated by modeling \cite{liao2020spin}, and for this reason we have not yet attempted to make prototype nanoscale devices for testing. It is important to distinguish our goal, efficient switching of nanoscale PMA samples, from previous demonstrations of switching in micron-scale samples. Many studies have shown that even very weak out-of-plane torques can assist spin-orbit-torque switching of micron-scale PMA samples, but in all such studies performed to date the critical current is still determined primarily by the conventional in-plane component of spin-orbit torque that acts to shift domain walls, with the out-of-plane component acting only to provide a bit of symmetry breaking required to make the switching ``field free".  Studies of spin-orbit switching of micron-scale samples via domain-wall motion are therefore not directly applicable to the development of efficient switching of practical nanoscale PMA devices driven primarily by the out-of-plane anti-damping torque because the switching mechanisms are not the same \cite{zhang2015spin}. \\

    In summary, we have demonstrated a strategy for achieving a partly-canted magnetization configuration in a magnetic spin-source layer at zero applied magnetic field, and have shown that this configuration allows for the generation of vertically-flowing spin current with an out-of-plane spin component, which can therefore exert an out-of-plane anti-damping spin-orbit torque on an adjacent magnetic layer.  The use of the ferromagnetic moment to provide the symmetry-breaking needed to enable an out-of-plane anti-damping torque circumvents the challenges posed by low-crystalline-symmetry materials and antiferromagnets for switching magnetic memories with perpendicular magnetic anisotropy.  Based on the dependence of the out-of-plane torques on the angle of the applied electric field relative to the canted magnetization, the out-of-plane anti-damping torques generated by SrRuO$_3$ can be well-described as due to the sum of a spin AHE and a PHE spin current.  The maximum out-of-plane torque efficiency that we measure, $\xi^Z_{DL} = 0.020 \pm 0.002$ is still below the threshold needed for spin-orbit-torque MRAM to achieve better energy efficiency than conventional spin-transfer-torque MRAM.  (Extended Data Table ~\ref{table:1} presents a list of  materials systems in which out-of-plane anti-damping torques have been measured quantitatively.)  Nonetheless, our study opens the door to explore torque generation by other magnetic materials in which magnetocrystalline anisotropy might be used to stabilize similar partly-canted magnetization configurations \cite{belashchenko2024exchange}. 
    %spin currents produced in ferromagnet with canted magnetization, as we showed that in principle the orientation of magnetic moment of a ferromagnet with relatively large magnetocrystalline anisotropy can be oriented favorably to produce more possible orientations of spins. This mechanism holds greater promise for technological application than low-crystalline and antiferromagnetic materials as it relaxes the requirement of single-crystallinity and requires smaller magnetic field to magnetize the as-grown multi-domain state.
    
\clearpage  
\printbibliography
\noindent 
\textbf{Methods}
\vspace{5mm}
\\
\noindent\textbf{Sample preparation}\\
SrRuO$_3$ thin films were grown with pulsed laser deposition (PLD). The thin films were grown on slightly miscut $(001)$-oriented SrTiO$_3$ substrates at 700 $^\circ$C, 100 mTorr oxygen partial pressure, and a laser energy of 1.5 J/cm$^2$ at 2 Hz. After growth, the samples were cooled down at 5 $^\circ$C/min to room temperature at an atmospheric oxygen pressure. The detection magnet layer (Py) was deposited in an AJA magnetron sputterer system with a base pressure of 2$\times$10$^{-8}$ Torr. A 1.5 nm Ta capping layer was then deposited on the Py to inhibit oxidation. 
\vspace{5mm}
\\
\noindent\textbf{Device fabrication}\\
ST-FMR and Hall devices with various $\psi$ angles to the crystal axis were patterned using a Heidelberg laser direct writer. The SrRuO$_3$ layer was etched with NaIO$_4$ solution (0.1 mol/L) and other layers were etched by ion milling. Contacts made from Pt (100 nm)/Ti (5 nm) were sputtered in an AJA chamber and then lifted off. For ST-FMR devices, strips with dimension 10 $\mu$m $\times$ 40 $\mu$m were in contact with ground-signal-ground electrodes. For Hall devices, the dimensions of the Hall bar were 20 $\mu$m $\times$ 50 $\mu$m.
\vspace{5mm}
\\
\noindent\textbf{Magnetoresistance measurement}\\
Hall devices with various in-plane orientations ($\psi$) were patterned on SrRuO$_3$/Py samples. A direct current of 100 $\mu$A was applied with Keithley 2400 d.c.\ current source meter in the current channel and a longitudinal voltage was measured with Keithley 2181A nanovoltemeter. For the magnetic-field-angle dependent magnetoresistance measurements, an external magnetic field was rotated in a plane perpendicular to the applied current. The measurements were repeated on devices with in-plane orientation $\psi$ from $0^\circ$ to $90^\circ$ at every $15^\circ$ to identify both the magnet canting direction and canting angle. To investigate the exchange coupling between SrRuO$_3$ and Py, the external magnetic field was swept in the plane from positive to negative along the current direction, while recording the longitudinal voltage. The measurement was repeated on devices oriented along both the $[100]_{pc}$ and $[010]_{pc}$ directions.
\vspace{5mm}
\\
\noindent\textbf{ST-FMR measurement}\\
ST-FMR devices with ground-signal-ground (GSG) contacts were fabricated from the SrRuO$_3$/Py bilayers. A microwave current at a fixed frequency (5 GHz - 5.5 GHz) and power was applied. A dc mixing voltage was measured with a DSP lockin amplifier while sweeping the external magnetic field from positive to negative at a fixed angle ($\varphi$) to the applied current. The same ST-FMR measurement was repeated at magnetic field angles $\varphi$ from $0^\circ$ to $170^\circ$ at every $10^\circ$. This magnetic-field-angle-dependent ST-FMR measurement was performed on devices with applied-current angles $\psi$ (in the film plane relative to the $[100]_{pc}$ direction)  from $0^\circ$ to $165^\circ$ in $15^\circ$ increments. The temperature-dependent ST-FMR measurements were carried out in a home-built cryostat cooled by liquid nitrogen. 
\vspace{5mm}
\\
\noindent\textbf{Acknowledgements}\\
X.H.\ was supported by the US Department of Energy (DOE),  Office of Basic Energy Sciences (BES) under award number DE-SC0017671. D.P., D.G.S., and D.C.R.\ acknowledge support from SUPREME, one of seven centers in JUMP 2.0, a Semiconductor Research Corporation program sponsored by DARPA. Z.T.\ acknowledges that this work is supported by the Air Force Office of Scientific Research under award number FA9550-24-1-0266. Any opinions, findings, and conclusions or recommendations expressed in this material are those of the author(s) and do not necessarily reflect the views of the United States Air Force. L.W.M.\ acknowledges support from the the National Science Foundation under Grant DMR-2329111. T.M.J.C.\ and M.O.\ were supported in part by the US National Science Foundation (NSF) under DMR-2104268. C.B.E.\ acknowledges support from Vannevar Bush Faculty Fellowship (ONR N00014-20-1-2844), and the Gordon and Betty Moore Foundation’s EPiQS Initiative, Grant No. GBMF9065. Thin Film Synthesis at the University of Wisconsin–Madison was supported by the U.S. Department of Energy, Office of Science, Office of Basic Energy Sciences, under Award No. DE-FG02-06ER46327. The devices were fabricated using the shared facilities of the Cornell NanoScale Facility, a member of the National Nanotechnology Coordinated Infrastructure (supported by the NSF via grant NNCI-2025233) and the facilities of the Cornell Center for Materials research.
\vspace{5mm}
\\
\noindent\textbf{Author contributions}\\
X.H.\ and D.C.R.\ conceived and designed the research. Z.T., Y.L., K.L., and C.W.\ carried out the oxide thin film growth. X.H. and M.O. performed the magnetron sputtering. X.H.\ and T.M.J.C.\ carried out the low temperature ST-FMR measurements with help from D.A.P. X.H.\ did the device fabrication with help from D.A.P. X.H.\ and D.A.P.\ performed the angle-dependent magnetoresistance measurements. T.M.J.C.\ simulated the angular-dependent magnetoresistance. H.Z.\ and D.F.S.\ performed the phenomenological symmetry analysis. Y.L., D.Y., C.B.E., D.G.S., and L.W.M.\ gave suggestions on the experiments. All authors discussed the results and assisted in the preparation of the manuscript. 
\vspace{5mm}
\\
\noindent\textbf{Competing interests}\\
The authors declare no competing interests
\vspace{5mm}
\\
\noindent\textbf{Additional information}\\ Correspondence and requests for materials should be addressed to X.H. and D.C.R.\\
\newpage
    \begin{figure}[t!]
    	\centering
    	\includegraphics[width=0.8\textwidth]{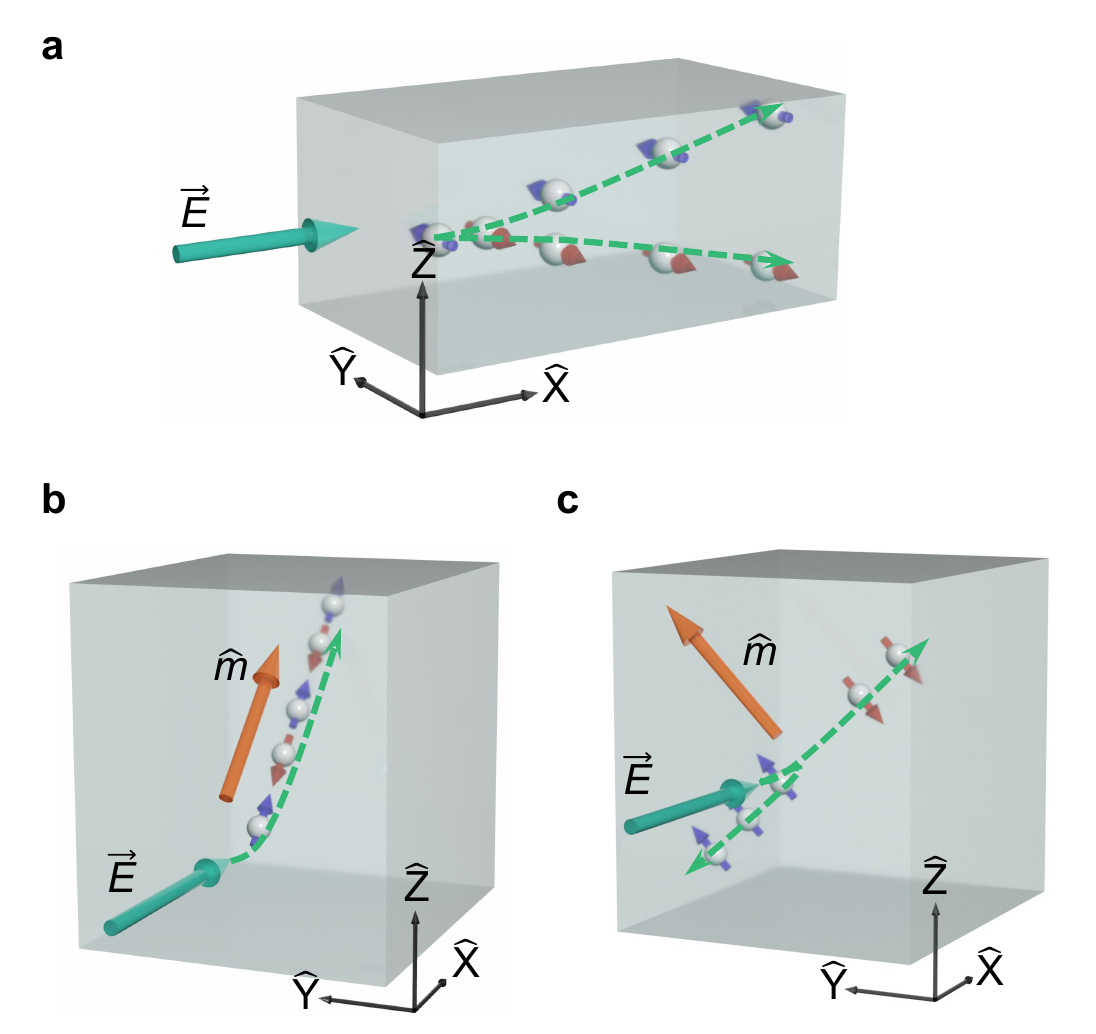}
    	\caption{\textbf{Tilted spin currents produced by a canted magnetization. a, }
        Illustration of the spin current allowed by symmetry for a non-magnetic material with high crystal symmetry, for which a spin current flowing in the $\hat{Z}$ direction must have spin polarization oriented in the $\pm \hat{Y}$ direction. \textbf{b, }Illustration of a tilted spin current originating from the planar Hall effect in a ferromagnet, relevant for the case in which an electric field ($\vec{E}$) is applied along the $\hat{X}$ direction and the magnetic moment is canted away from the $\hat{Z}$ direction into the $\hat{X}$ direction. \textbf{c, }Illustration of a tilted spin current originating from the spin anomalous Hall effect in a ferromagnet, relevant for the case in which the magnetic moment is canted away from the $\hat{Z}$ direction into the $\hat{Y}$ direction.}
    	\label{fig:1}
    	%\vspace{80pt}
    \end{figure}
    \newpage
    \begin{figure}[t!]
    	\centering
    	\includegraphics[width=1\textwidth]{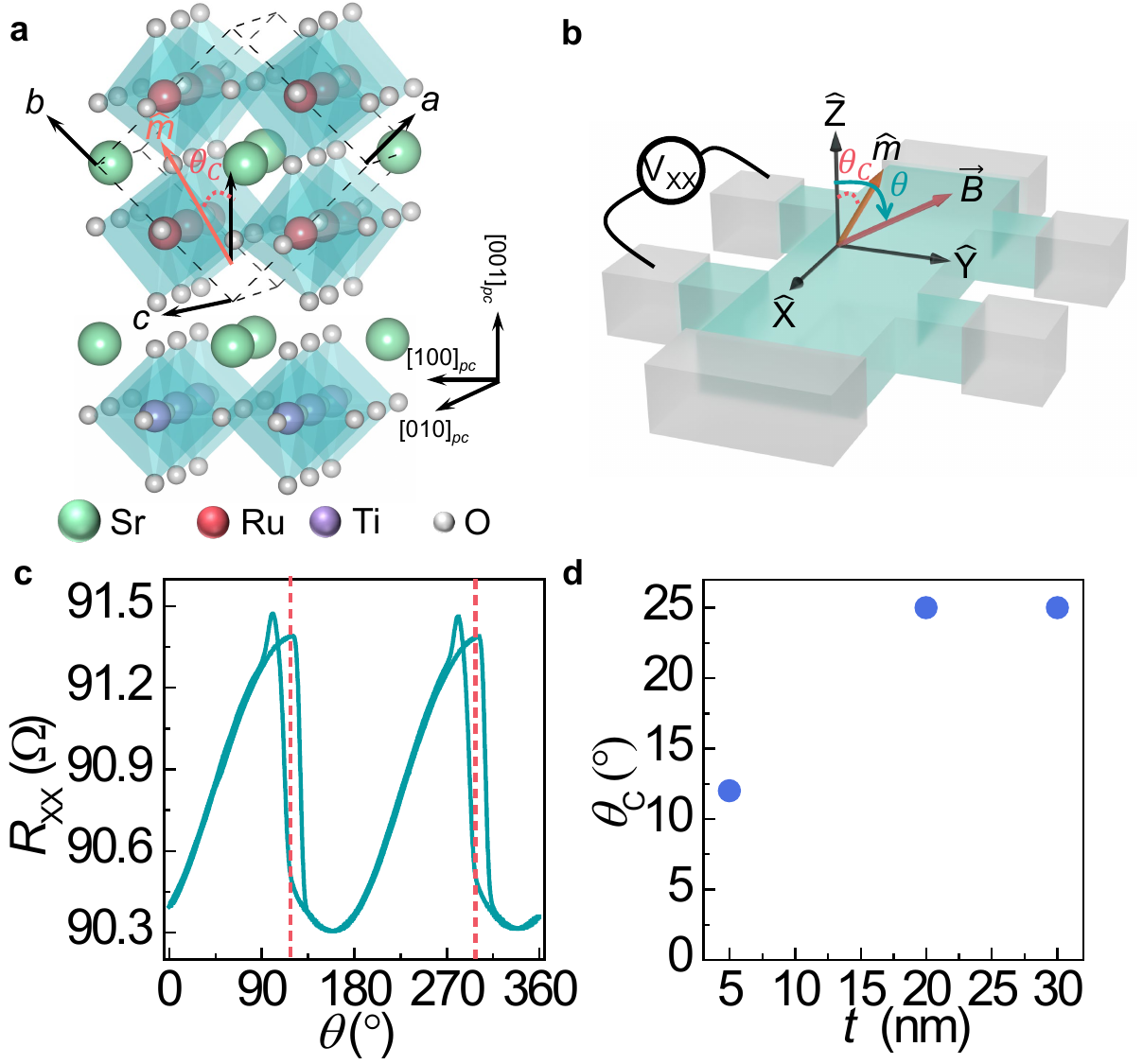}
    	\caption{\textbf{Magnetic anisotropy of SrRuO$_3$ thin films. a, } A side view showing the epitaxial relationship of SrRuO$_3$ grown on an $(001)$-oriented SrTiO$_3$ substrate, where the $[010]_{pc}$ axis points out of the page. $a$, $b$, and $c$ are the orthorhombic unit cell axes of SrRuO$_3$. In thin films, the direction of the magnetic easy axis ($\theta_C$) can differ from the $b$ axis.  \textbf{b, }Device geometry for the angular-dependent magnetoresistance measurements. A longitudinal voltage ($V_{XX}$) is recorded while rotating the external magnetic field ($\vec{B}$) in a plane perpendicular to the applied current.  $\theta$ is the external magnetic field angle from the surface normal ($[001]_{pc}$). \textbf{c, }An example angular dependent magnetoresistance measurement at 10 K, where longitudinal resistance ($R_{XX}$) is plotted as a function of magnetic field angle $\theta$ for a sample SrRuO$_3$(20 nm)/Py (5 nm). A d.c.\ current of 100 $\mu$A is applied along the $[010]_{pc}$ orientation and a magnetic field 6 T is rotated in the $(010)_{pc}$ plane. \textbf{d, }Summary of the easy-axis tilt angle ($\theta_C$) for SrRuO$_3$ films of various thicknesses ($t$) at 10 K.  
         }
    	\label{fig:2}
    	%\vspace{80pt}
    \end{figure}
    \newpage
    \begin{figure}[t!]
    	\centering
    	\includegraphics[width=1\textwidth]{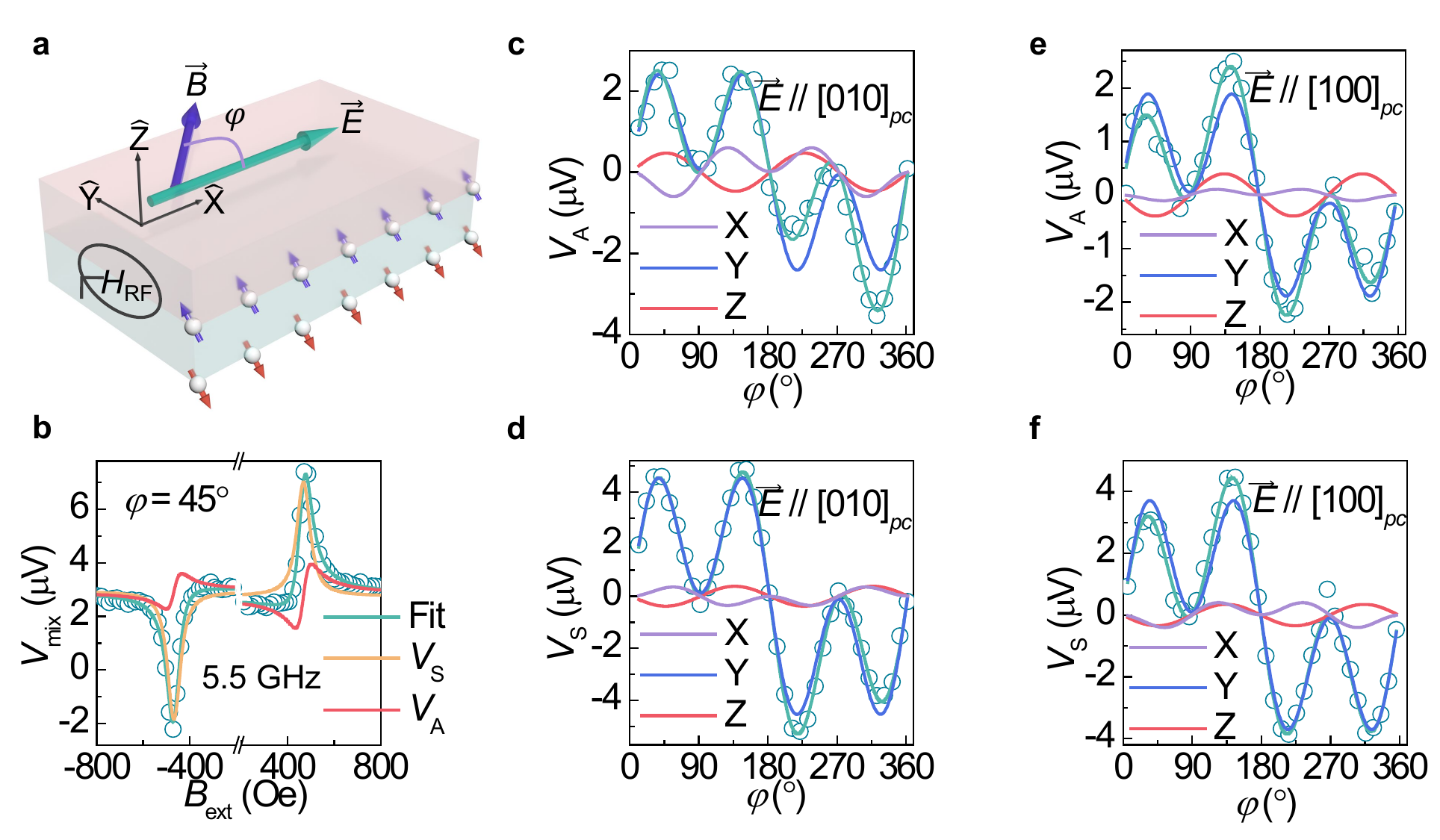}
    	\caption{\textbf{Spin torques characterized by spin-torque ferromagnetic resonance. a, } A schematic of a SrRuO$_3$ (5 nm)/Py (5 nm) bilayer  used in the ST-FMR measurements. Red layer: Py; Teal layer: SrRuO$_3$; $H_{RF}$: Oersted field produced by the applied microwave current; $\vec{B}$: external magnetic field; $\vec{E}$: electric field; $\varphi$: angle between external magnetic field and microwave current. \textbf{b, } Example ST-FMR spectrum, with the d.c.\ mixing voltage ($V_{mix}$)  plotted as a function of external magnetic field ($B$). In this example, the ST-FMR measurement is done at 110 $K$ with $E$ along [010]$_{pc}$ and $\varphi = 45^\circ$. The red and orange curves represent the anti-symmetric ($V_A$) and symmetric ($V_S$) components of the mixing voltage respectively. \textbf{c, } Anti-symmetric mixing voltage as a function of magnetic field angle. X (purple), Y (blue), Z (red) components of the anti-symmetric mixing voltages are determined as the fit components proportional to $\sin(2\varphi)\sin\varphi$, $\sin(2\varphi)\cos\varphi$ and $\sin(2\varphi)$ respectively. \textbf{d, } Symmetric mixing voltage as a function of magnetic field angle. X (purple), Y (blue) and Z (red) components of the symmetric mixing voltages are determined as the fit components proportional to $\sin(2\varphi)\cos\varphi$, $\sin(2\varphi)\cos\varphi$ and $\sin(2\varphi)$ respectively. The electric field is applied approximately along the $[010]_{pc}$ crystallographic axis in the plane for \textbf{c} and \textbf{d}. \textbf{e, f } Anti-symmetric and symmetric mixing voltages as a function of magnetic field angle for an electric field applied approximately along the $[100]_{pc}$ crystallographic axis.
         }
    	\label{fig:3}
    	%\vspace{80pt}
    \end{figure}
    
    \begin{figure}[t!]
    	\centering
    	\includegraphics[width=1\textwidth]{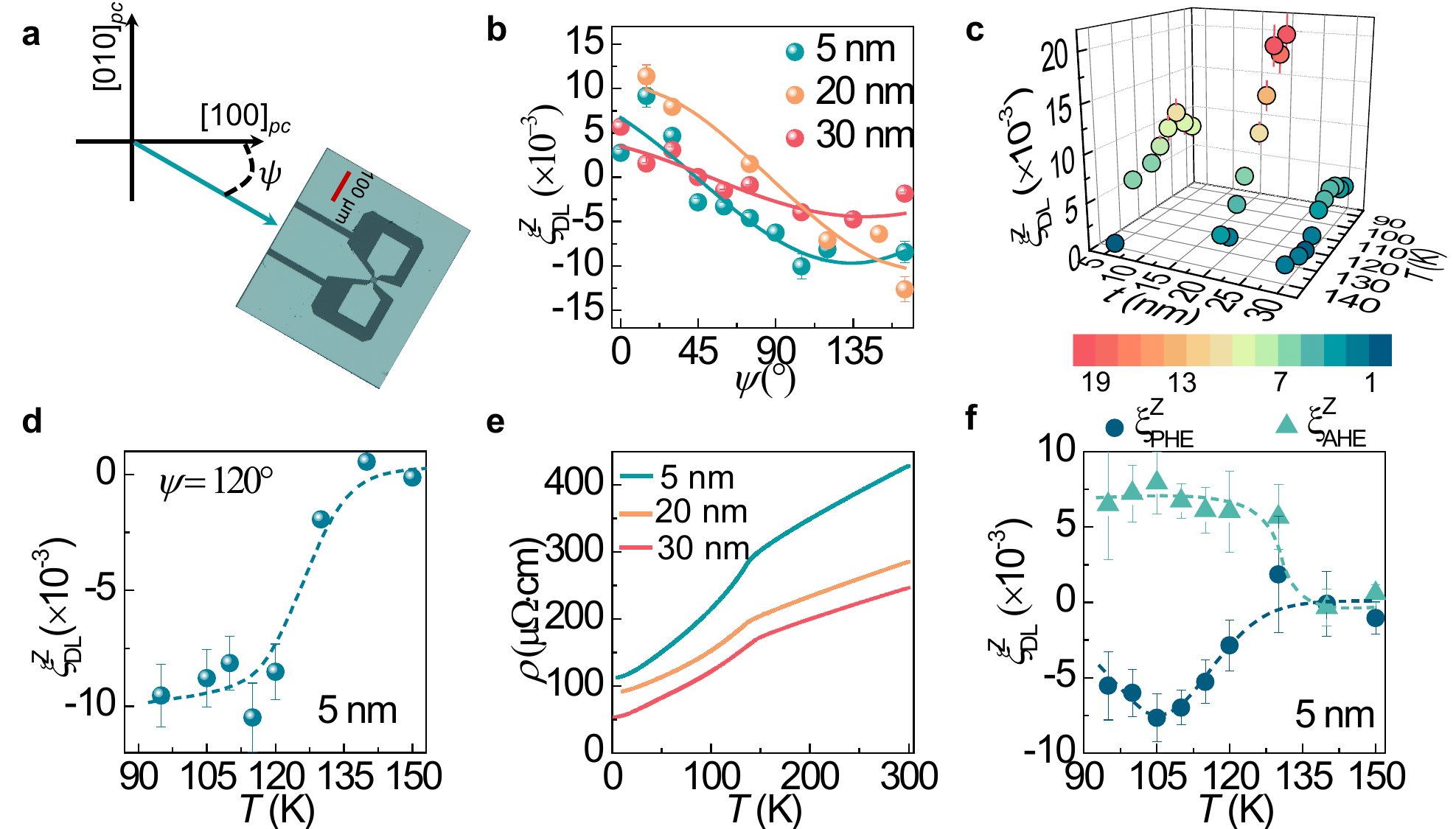}
    	\caption{\textbf{Dependence of out-of-plane anti-damping torque on the orientation of the applied electric field and temperature. a, } A schematic of the ST-FMR measurement geometry, for $\vec{E}$ is oriented at a $\psi$ angle to [100]$_{pc}$. \textbf{b, } Out-of-plane anti-damping torque efficiency as a function of $\psi$ for a SrRuO$_3$ (5 nm)/Py (5 nm) sample (110 K), a SrRuO$_3$ (20 nm)/Py (5 nm) sample (110 K), and a SrRuO$_3$ (30 nm)/Py (5 nm) sample (105 K). The solid curves represent fits to the data as a linear superposition of $\sin\psi$ and $\cos\psi$, used to determine the AHE and PHE contributions as in panel \textbf{f} below. \textbf{c, }Temperature dependent out-of-plane damping-like torque efficiency for SrRuO$_3$ of 5 nm, 20 nm and 30 nm. \textbf{d, }Representative temperature dependent out-of-plane anti-damping torque efficiency at $\psi=120^\circ$ for the SrRuO$_3$ (5 nm)/Py (5 nm) sample. The dashed line is a guide to the eye. \textbf{e, }Resistivity as a function of temperature for SrRuO$_3$ of various thicknesses. \textbf{f, }The PHE ($\xi_{PHE}^Z$) and AHE ($\xi_{AHE}^Z$) components of the out-of-plane anti-damping torque efficiencies as a function of temperature for $5$ nm SrRuO$_3$. The dashed lines are a guide to the eye. Error bars in \textbf{b, c, d} are dominated by the uncertainties of film thicknesses, and error bars in \textbf{f} represent fitting uncertainties. 
         }
    	\label{fig:4}
    	%\vspace{80pt}
    \end{figure}
\newpage
\setcounter{figure}{0}
\captionsetup[figure]{labelfont={bf}, name={Extended Data Fig.}, labelsep=period}
\begin{figure}[t!]
    	\centering
    	\includegraphics[width=1\textwidth]{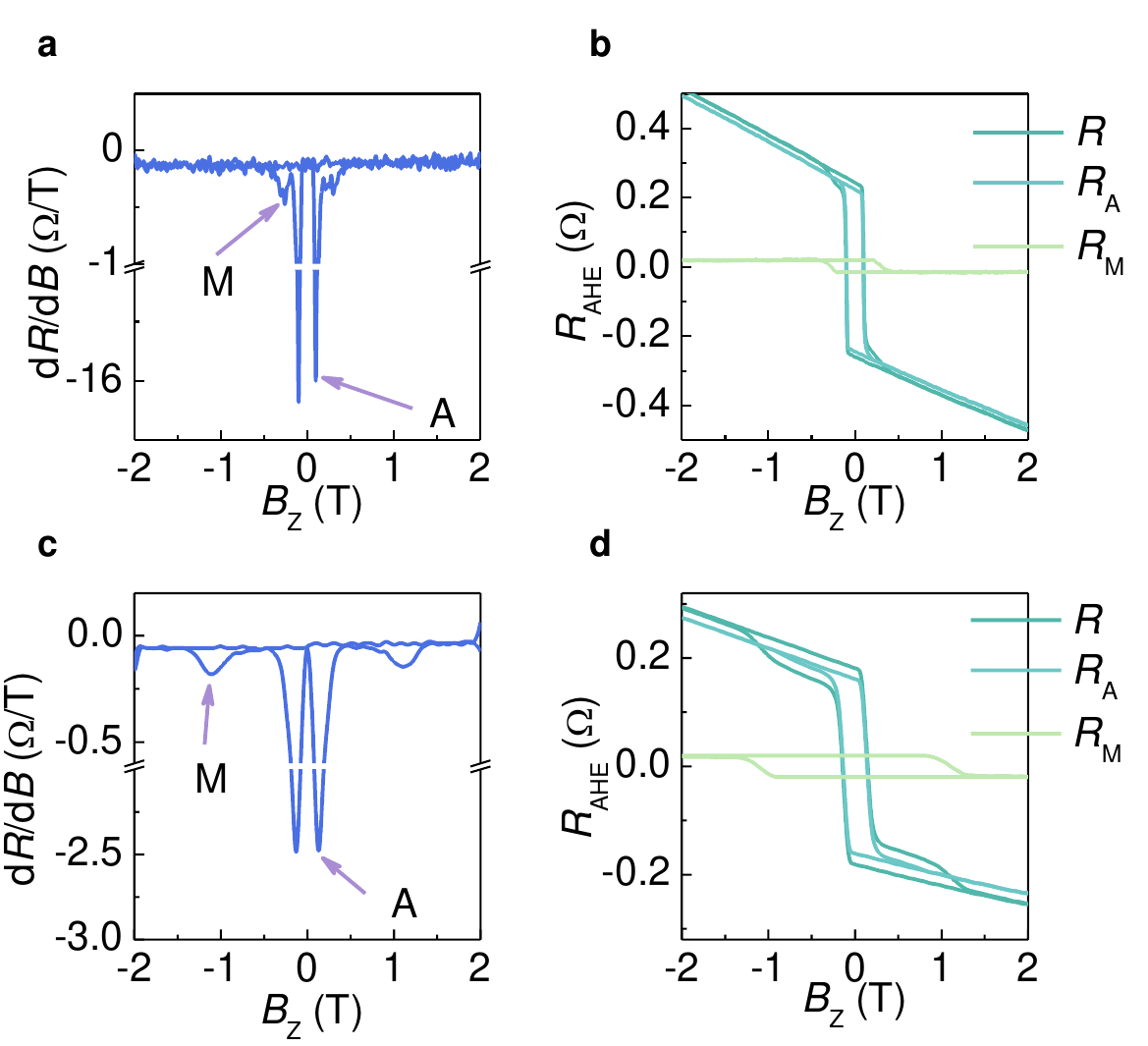}
    	\caption{\textbf{Domain structure characterized using anomalous Hall effect measurements. a, } The first derivative of anomalous Hall resistance as a function of external magnetic field for a single-layer $5$ nm SrRuO$_3$ thin film at $110$ K. `A' denotes the dominant domain and `M' denotes minor domains present in the sample. \textbf{b, } Anomalous Hall resistance for the same single-layer $5$ nm SrRuO$_3$ thin film at $110$ K. $R_A$ and $R_M$ are the contributions from the `A' domain and the minor domain to the anomalous Hall resistance. \textbf{c, } The first derivative of anomalous Hall resistance as a function of external magnetic field for a $10$ nm SrRuO$_3$ thin film at $110$ K. \textbf{d, } Anomalous Hall resistance for the same $10$ nm SrRuO$_3$ film at $110$ K. 
         }
    	\label{fig:efig_1}
    	%\vspace{80pt}
    \end{figure}
\newpage
\begin{figure}[t!]
    	\centering
    	\includegraphics[width=1\textwidth]{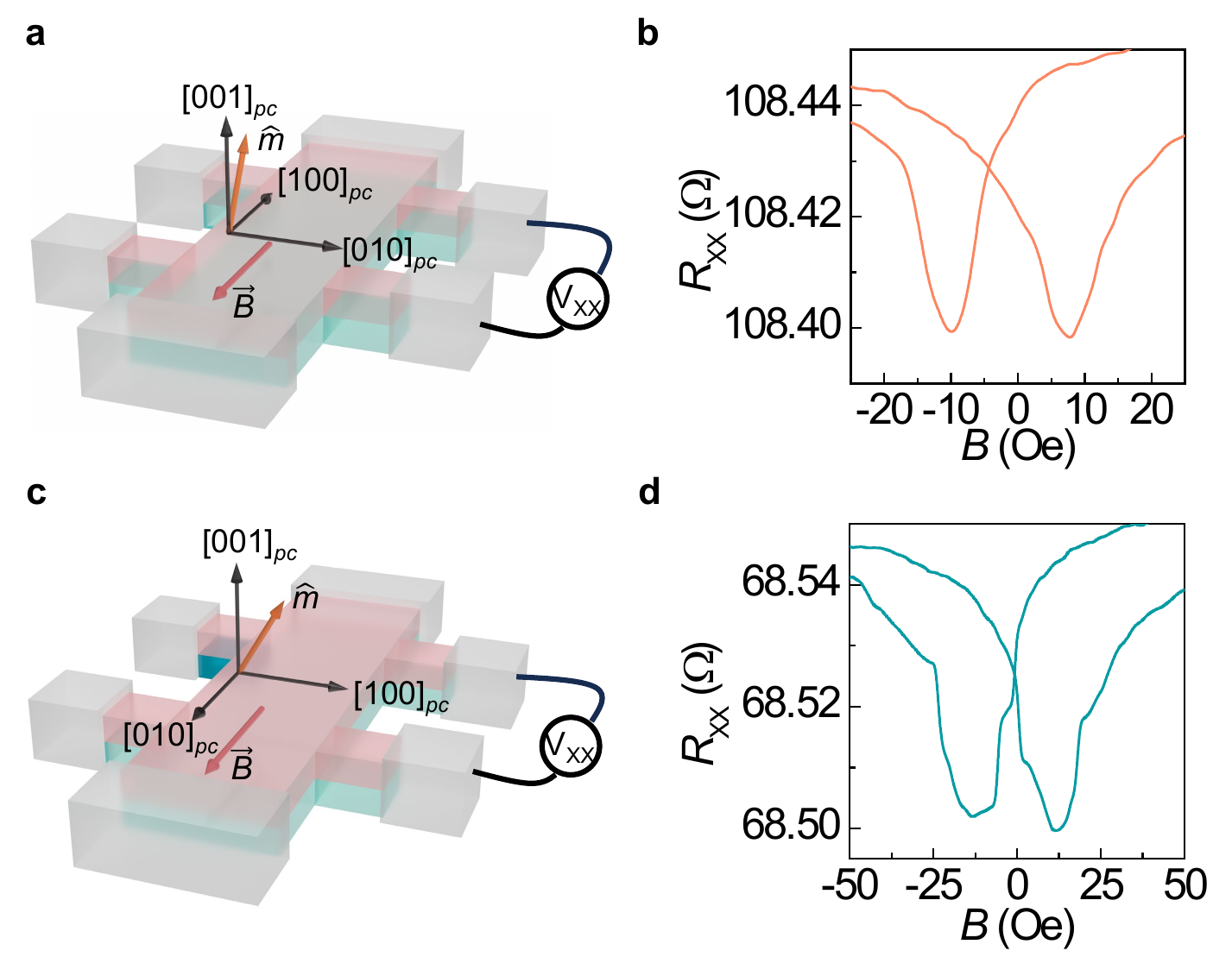}
    	\caption{\textbf{A check for exchange coupling between SrRuO$_3$ (10 nm) and Py (5 nm). a, }Schematic illustration of the measurement geometry to test for exchange coupling along the [100]$_{pc}$ direction. Longitudinal voltage ($V_{XX}$) is measured while an external magnetic field is swept along the [100]$_{pc}$ direction. \textbf{b, }Measured longitudinal resistance as a function of external magnetic field at $80$ K for the geometry in panel \textbf{a}. Two dips are observed, at $8$ Oe and $-10$ Oe, indicating the absence of any significant exchange bias along the [100]$_{pc}$ direction. \textbf{c, }Measurement geometry to test for exchange coupling along the [010]$_{pc}$ direction. \textbf{d, }Measured longitudinal resistance as a function of external magnetic field at $80$ K for the geometry in panel \textbf{c}. Two dips are observed symmetrically located at $\pm 13$ Oe, indicating the absence of exchange bias along [010]$_{pc}$ direction.}
    	\label{fig:efig_2}
    	%\vspace{80pt}
\end{figure}
\newpage
\begin{figure}[t!]
    	\centering
    	\includegraphics[width=1\textwidth]{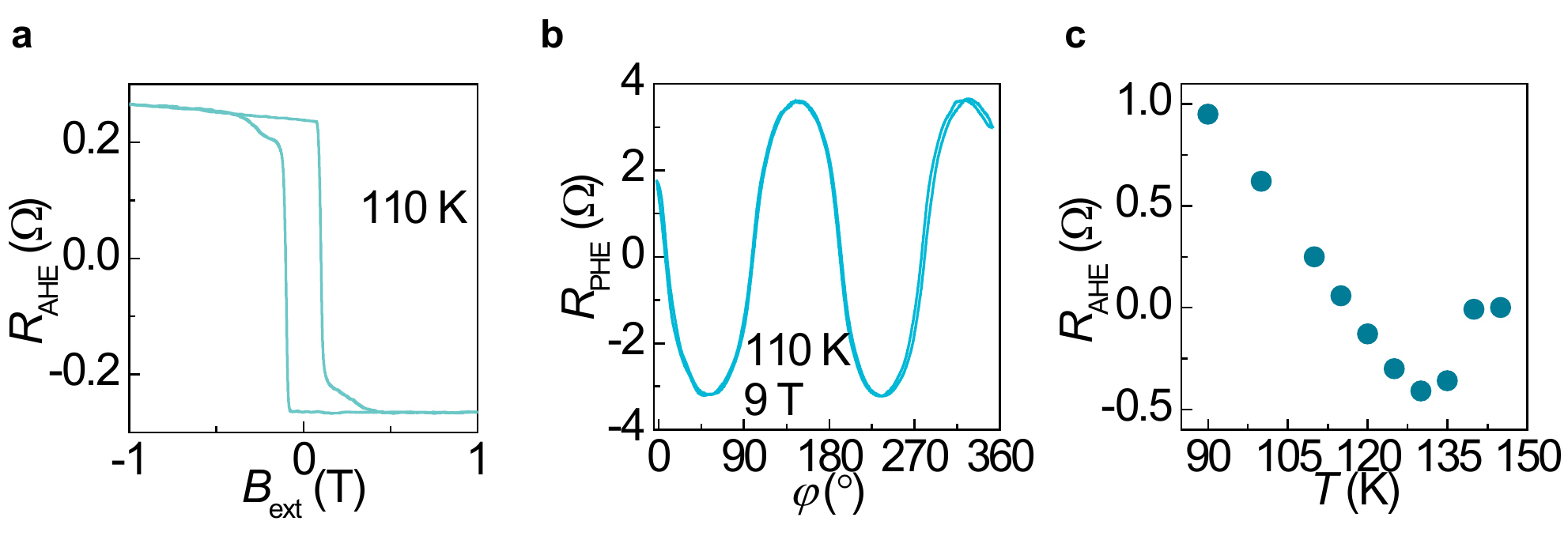}
    	\caption{\textbf{Anomalous Hall and planar Hall effects. a, }Anomalous Hall resistance for a $5$ nm SrRuO$_3$ film at $110$ K. \textbf{b, }Planar Hall effect for the same $5$ nm SrRuO$_3$ sample at $110$ K. An external magnetic field of $9$ T is applied. \textbf{c, }Anomalous Hall resistance as a function of temperature for the $5$ nm SrRuO$_3$ film.
         }
    	\label{fig:efig_3}
    	%\vspace{80pt}
\end{figure}  
%\clearpage

\begin{figure}[t!]
    	\centering
    	\includegraphics[width=1\textwidth]{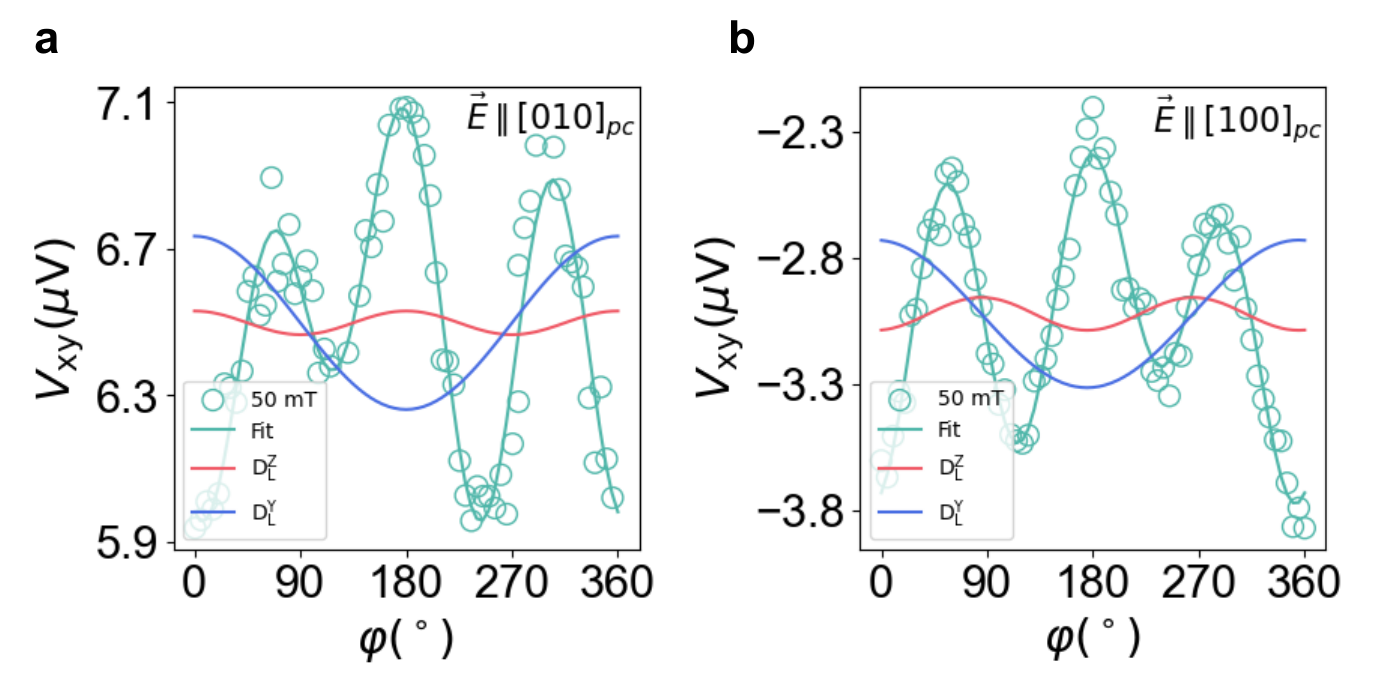}
    	\caption{\textbf{Second harmonic Hall voltage measurements for the sample SrRuO$_3$ (5 nm)/Py (5 nm) at 100 K. a, b, }Second harmonic Hall voltages ($V_{xy}$) as a function of the in-plane orientation ($\varphi$) of a magnetic field of 50 mT for electric field ($\vec{E}$) along [010]$_{pc}$ and [100]$_{pc}$ respectively.}
    	\label{fig:efig_4}
    	%\vspace{80pt}
\end{figure} 

\begin{figure}[t!]
    	\centering
    	\includegraphics[width=1\textwidth]{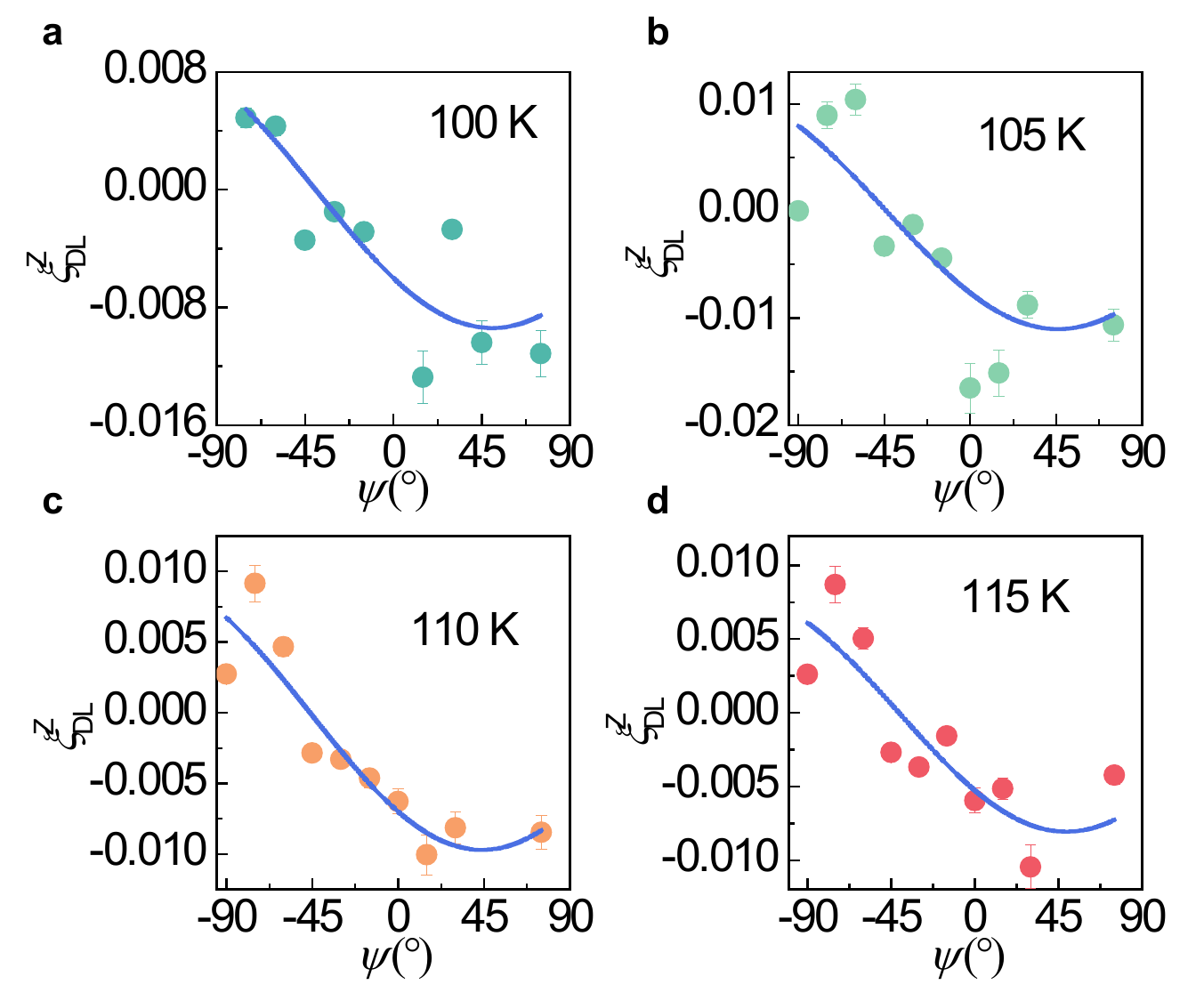}
    	\caption{\textbf{Measured out-of-plane anti-damping torque efficiencies at various temperatures. a, b, c, d, }Out-of-plane damping-like torque efficiency as a function of the angle $\psi$ of the applied electric field at 100 K, 105 K, 110 K and 115 K, respectively. The curves in blue are fits to the data of the form $\xi_{DL}^Z =\xi_{AHE}^Z\sin\psi +\xi_{PHE}^Z\cos\psi$. Error bars are calculated based on the uncertainties of film thicknesses.}
    	\label{fig:efig_5}
    	%\vspace{80pt}
\end{figure} 

\begin{figure}[t!]
    	\centering
    	\includegraphics[width=1\textwidth]{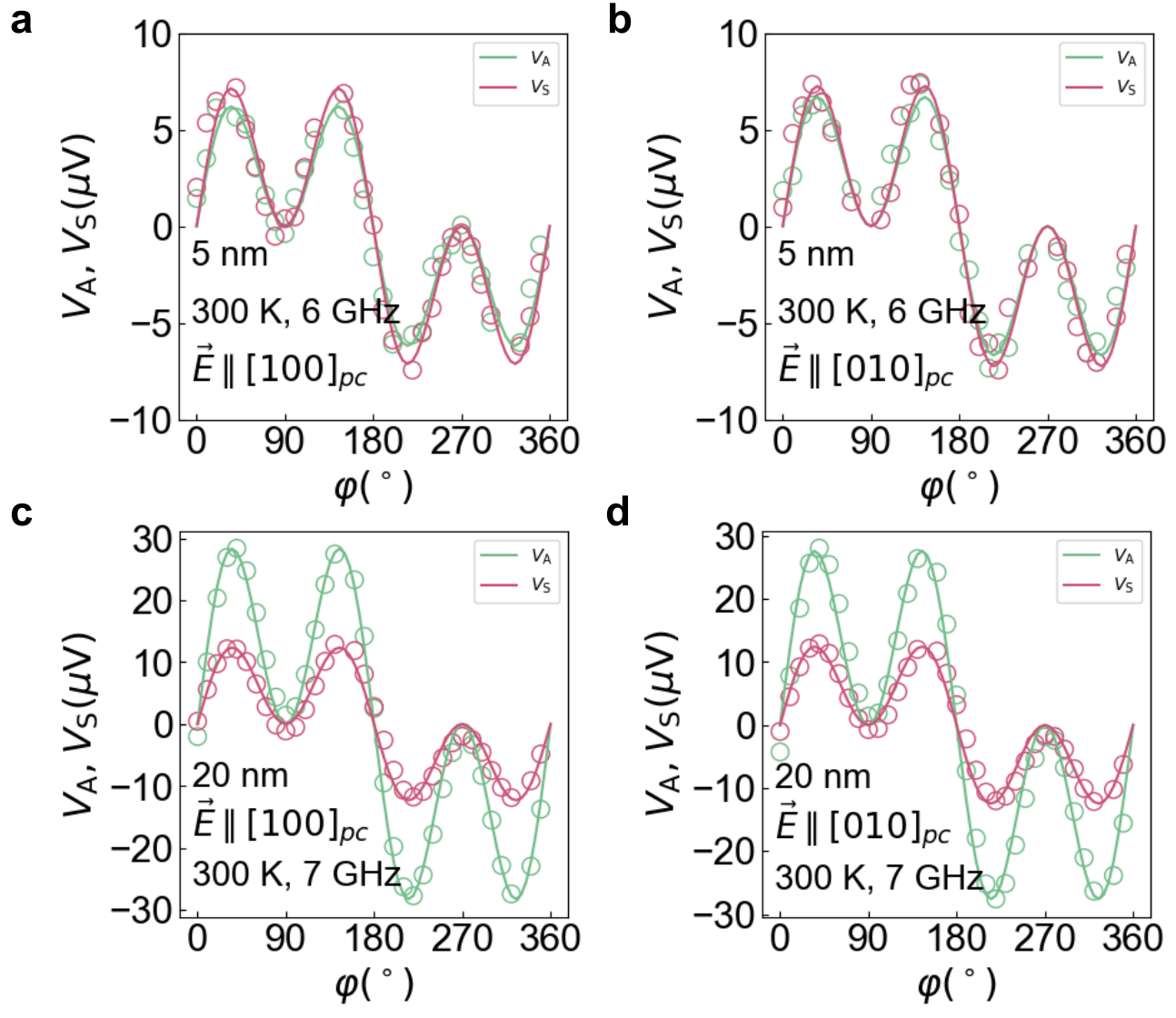}
    	\caption{\textbf{ST-FMR measurements at room temperature for samples SrRuO$_3$ (5, 20 nm)/Py (5 nm). a, b, }Anti-symmetric and symmetric components of the mixing voltage for electric fields applied along [100]$_{pc}$ and [010]$_{pc}$ for SrRuO$_3$ (5 nm)/Py (5 nm). The solid curve is a fit to Y component, proportional to $\sin(2\varphi)\cos(\varphi)$. \textbf{c, d, }Anti-symmetric and symmetric components of the mixing voltage for electric fields applied along [100]$_{pc}$ and [010]$_{pc}$ for SrRuO$_3$ (20 nm)/Py (5 nm). The solid curve is a fit to Y component, proportional to $\sin(2\varphi)\cos(\varphi)$}
    	\label{fig:efig_6}
    	%\vspace{80pt}
\end{figure} 

\begin{figure}[t!]
    	\centering
    	\includegraphics[width=1\textwidth]{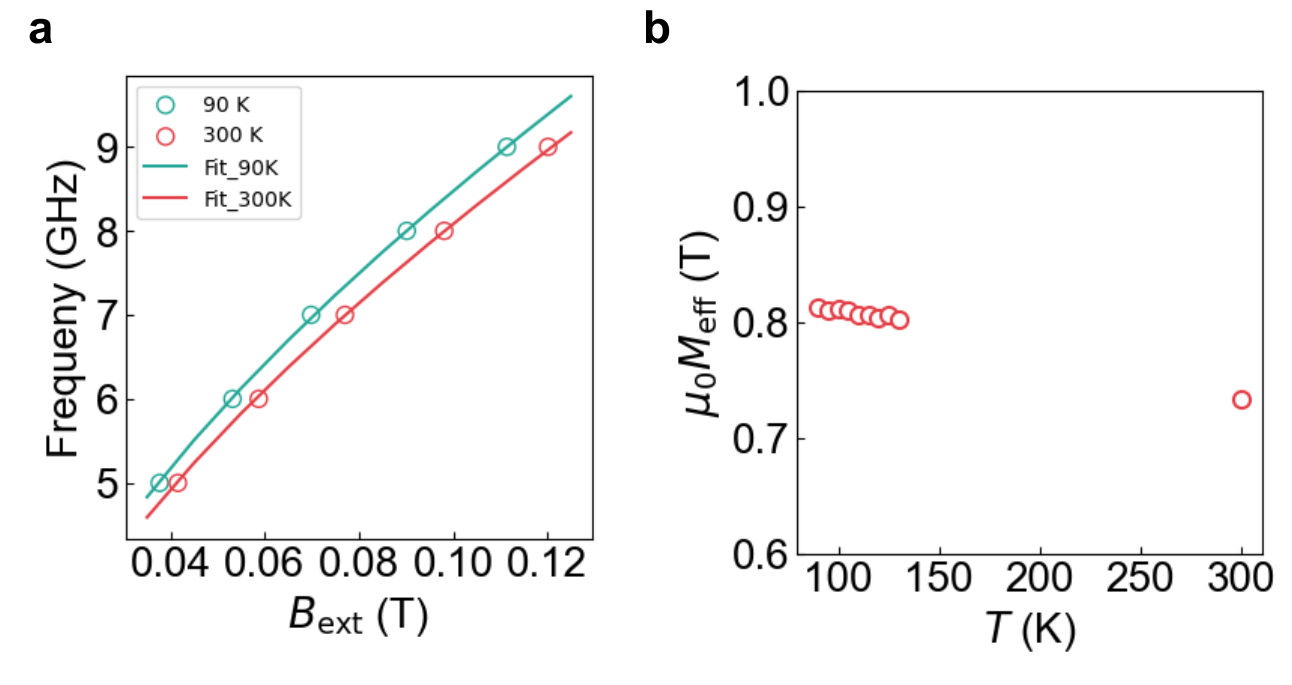}
    	\caption{\textbf{Effective magnetization analysis for sample SrRuO$_3$ (5 nm)/Py (5 nm). a, }Measured resonant magnetic fields as a function of frequency at 300 K and 90 K. The solid curves are Kittel fits (to $f=\gamma/2\pi \sqrt{B_{ext}(B_{ext}+\mu_0M_{eff})}$). \textbf{b, }Effective magnetization for SrRuO$_3$ (5 nm)/Py (5 nm) as a function of temperature.}
    	\label{fig:efig_7}
    	%\vspace{80pt}
\end{figure} 
\newpage
        
\begin{table}[hbt!]
\centering
\begin{tabular}{ |c|c|c|c|c|} 
 \hline
 Material & $\xi_{DL}^Z$ & $\sigma_{SH}^Z (\Omega$m)$^{-1} $ & $\rho (\mu \Omega$cm) & ref\\ 
 \hline
 \multicolumn{5}{|c|}{Low crystal symmetry}\\
 \hline
 WTe$_2$ & 0.014&$3.6\times 10^3$ & 380  & \cite{macneill2017control} \\ 
 \hline
 MoTe$_2$& 0.008&$1.4\times 10^3$&550&\cite{stiehl2019layer}\\
 \hline 
 (114) MnPd$_3$&0.008&$14\times 10^3$&60&\cite{dc2023observation}\\
 \hline
 PtTe$_2$/WTe$_2$&0.007-0.032&$(17-25)\times 10^3$&29-154&\cite{wang2024field}\\
 \hline
  \multicolumn{5}{|c|}{Antiferromagnets}\\
 \hline 
  RuO$_2$&0.008&$6\times 10^3$&140&\cite{bose2022tilted}\\
 \hline
 Mn$_3$GaN&0.019&$8.6\times 10^3$&220&\cite{nan2020controlling}\\
 \hline
 %(100) Mn$_3$Pt&0.03&$14\times 10^3$&220&\cite{bai2021control}\\
 %\hline
 \multicolumn{5}{|c|}{Ferromagnets}\\
 \hline
  $[001]_{pc}$ SrRuO$_3$ (20 nm)&0.02 (90 K)&$13.9\times10^3$&143.5&This work\\
 \hline
\end{tabular}
\caption{A survey of out-of-plane anti-damping torque in various systems for which quantitative measurements have been reported.  }
\label{table:1}
\end{table}
\end{document}